\documentclass[aps,prl,twocolumn,superscriptaddress]{revtex4}

\usepackage{amssymb}
\usepackage{graphicx}
\usepackage{amsmath}
\usepackage{epsfig}
\usepackage{graphicx}% Include figure files
\usepackage{dcolumn}% Align table columns on decimal point
\usepackage{bm}% bold math
\usepackage[export]{adjustbox}
\usepackage{soul,ulem}
\usepackage[usenames,dvipsnames]{color}%Use color package
\usepackage{braket}
\newcommand{\rev}[1]{{\color{black}#1}}

\begin{document}

\title{Unveiling the BEC-droplet crossover with Rayleigh superradiant scattering}% 

\author{Mithilesh K. Parit}
\thanks{These authors contributed equally.}
\affiliation{Department of Physics, The Hong Kong University of Science and Technology, Clear Water Bay, Kowloon, Hong Kong, China}

\author{Mingchen Huang}
\thanks{These authors contributed equally.}
\affiliation{Department of Physics, The Hong Kong University of Science and Technology, Clear Water Bay, Kowloon, Hong Kong, China}

\author{Ziting Chen}
\affiliation{Department of Physics, The Hong Kong University of Science and Technology, Clear Water Bay, Kowloon, Hong Kong, China}

\author{Yifei He}
\affiliation{Department of Physics, The Hong Kong University of Science and Technology, Clear Water Bay, Kowloon, Hong Kong, China}

\author{Haoting Zhen}
\affiliation{Department of Physics, The Hong Kong University of Science and Technology, Clear Water Bay, Kowloon, Hong Kong, China}
\affiliation{Department of Physics and Astronomy, Rice University, Houston, TX, USA}%

\author{Gyu-Boong Jo}
\email{gbjo@rice.edu}
\affiliation{Department of Physics and Astronomy, Rice University, Houston, TX, USA}%
\affiliation{Department of Electrical and Computer Engineering, Rice University, Houston, TX, USA}%
\affiliation{Smalley-Curl Institute, Rice University, Houston, TX, USA}%
\affiliation{Department of Physics, The Hong Kong University of Science and Technology, Clear Water Bay, Kowloon, Hong Kong, China}

\begin{abstract}
Light scattering plays an essential role in uncovering the properties of quantum states through light-matter interactions. Here, we explore the crossover from Bose-Einstein condensate (BEC) to droplets in a dipolar $^{166}$Er gas by employing superradiant light scattering as both a probing and controlling tool. We observe that the efficiency of superradiant scattering exhibits a non-monotonic behavior akin to the rate of sample expansion during the transition, signaling its sensitivity to the initial quantum state, and in turn, revealing the BEC-droplet crossover. Through controlled atom depletion via superradiance, we analyze the sample's expansion dynamics and aspect ratio to identify the BEC-droplet phases distinctly, supported by Gaussian variational ansatz calculations. Finally, using these two approaches, we track how the BEC-droplet crossover points shift under varying magnetic field orientations. Our results show that superradiance can provide a state-sensitive probe and implement rapid atom-number quenches, opening new avenues for studying quantum states in atomic matter.\end{abstract}

\maketitle

Light-matter interaction enables sensing, spectroscopy, quantum information processing, and materials characterization \cite{rev_lm1, MarkFox}. In ultracold atoms, it has broad applications---from realizing synthetic dimensions \cite{rev_lm2, rev_lm3, rev_lm4} and spin-orbit coupling \cite{rev_lm5, Zhang2018} to simulating condensed-matter systems \cite{rev_lm6} and imaging \cite{Img1, Img2, Img3}. Rayleigh superradiance scattering \cite{Dicke1954}, a light-matter interaction process commonly used in ultracold atom experiments, has been extensively studied across various systems \cite{SR_gen1, SR_gen2}. These include bosonic atom systems with isotropic s-wave interactions \cite{SR_BEC1, SR_BEC2, SR_BEC3, SR_BEC4, SR_PRA05, SR_PRL10, SR_PRA11, SR_PRL12, SR_PRA14, SR_PRA17}, atomic systems coupled to cavity modes \cite{SR_cavity07, SR_cavity10, SR_cavity14}, free fermions \cite{SR_Fermi}, and dipolar BECs with anisotropic dipolar interactions \cite{SR_dBEC}. However, most of previous works have focused on the fundamental collective enhancement from correlated multiple emitters. Here, we investigate superradiant scattering across the transition between a dipolar BEC and a macrodroplet, illustrating how the quantum state manifests and can be manipulated via light-matter interactions. We use the superradiant process as a state-sensitive observable influenced by the coherence of the initially prepared sample. Moreover, by adjusting its intensity of superradiant pulse, a well-defined fraction of atoms is coherently outcoupled on a timescale much faster than three-body decay, allowing the atom number to be tuned without driving the sample through a slow, lossy evolution.

In recent years, quantum droplets, the self-bound system of quantum gases \cite{petrov15}, have been observed in dipolar gases \cite{QD16_Pfau1, QD16_Pfau2, QD16_FF} as well as Bose-Bose mixtures \cite{QD18_BB1, QD18_BB2, QD18_BB3, Cavicchioli2025}, leading to the observation of supersolids \cite{ss1, ss2, ss3} with coherent multi-droplets. Both single macrodroplets and multidroplets have been observed where direction of magnetic field was along and perpendicular to elongated axis of the droplets, respectively \cite{rev_dp1}. While stripe states have been examined using tilted dipoles of dysprosium atoms~\cite{PRA17_Pfau} in an oblate trap, the formation of single macrodroplets with tilted dipoles and their coillective light-matter interactions \cite{qd_ex_sp} remain unexplored in an elongated trap. In this work, we investigate how the BEC-droplet transition points shift under varying magnetic field orientations using superradiance. Our work also complement the investigation of the excitation spectrum across the BEC-droplet crossover with Bragg spectroscopy \cite{Houwman2024}.

\begin{figure*}[ht]
	\includegraphics[width=\linewidth]{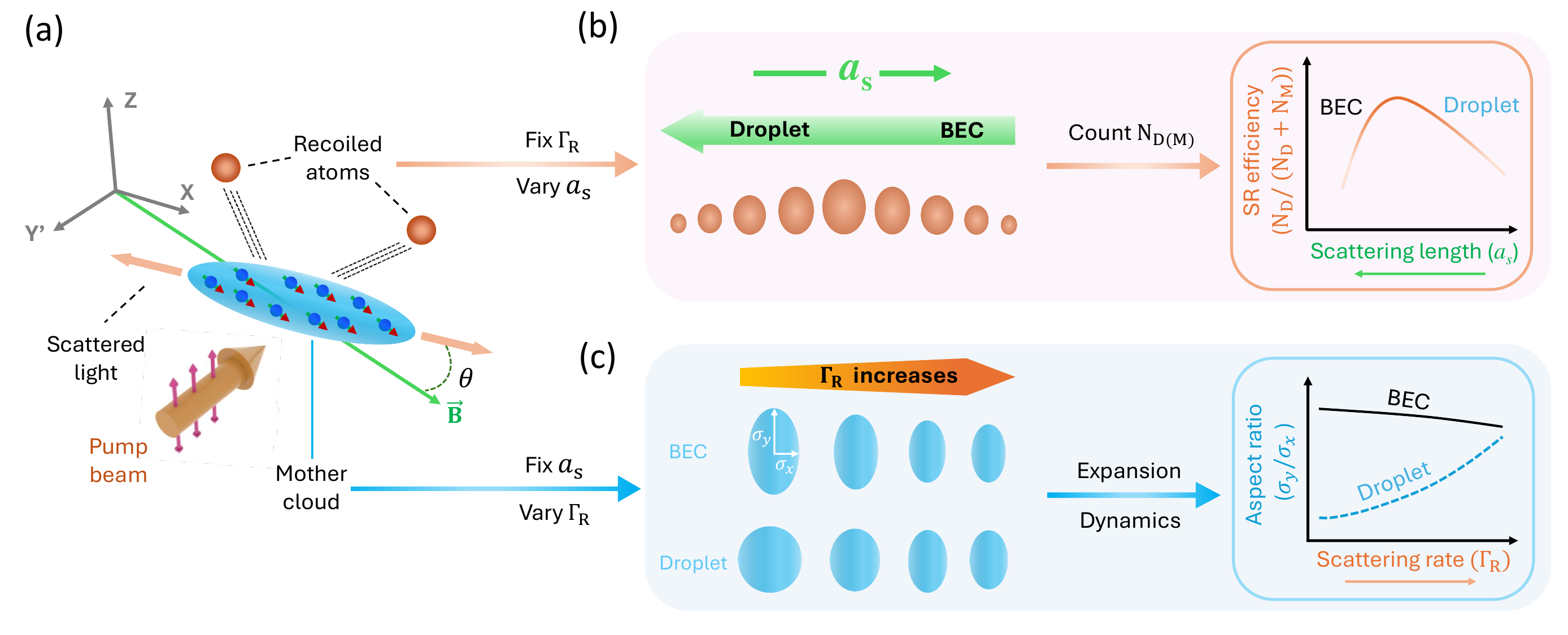}
	\caption{{\bf Superradiance from a BEC to a macrodroplet state.} (a) An elongated condensate is exposed to the superradiance beam linearly polarized along the $z$-direction, propagating along the $y-$axis. The magnitude ($B$) and direction ($\hat{r}(\theta)$) of the external magnetic field are independently controlled, leading to tunable $s$-wave scattering and dipolar interactions. (b) and (c) Schematic of mother cloud and recoiled atoms after 10 ms of free expansion. The scattering efficiency is enhanced near the crossover and suppressed in the deep droplet and BEC regimes. The aspect ratio of the mother cloud also exhibits distinct responses in the two regimes as the scattering rate $\Gamma_R$ is increased. The upper and lower panels summarize, respectively, the qualitative variation of superradiance efficiency across the crossover and the aspect-ratio response of the rapidly depleted mother cloud.}
	\label{fig1}
\end{figure*}

Our experiment illustrates how the superradiant process probes the quantum state of the sample as described in Fig.~\ref{fig1}. After superradiant scattering, two daughter clouds outcoupled from the original condensate (mother cloud) are formed after the expansion (Fig.~\ref{fig1}(a)). First, superradiant light scattering - which is sensitive to the sample's spatial coherence as well as its density and geometry - enables investigation of the BEC-droplet crossover (Fig.~\ref{fig1}(b)). In an elongated trap at finite temperature $T$, axial phase fluctuation primarily arises from low-energy excitation near $k_x\to 0$ \cite{Petrov01, PhaseFlucExp, Jo200732d}. This phase fluctuation occurs when the excitation wavelength becomes larger than the cloud's radial size, leading to 1D-like behavior \cite{Petrov01, PhaseFlucExp}, while density fluctuations maintain their 3D characteristics. With the decreasing chemical potential at constant $T$, fewer axial excitations become populated, leading to reduced phase fluctuation. As quantitatively demonstrated in our previous work \cite{SR_dBEC}, the superradiant threshold with the recoiled atom number $N$, described by $\dot{N}=(G-L)\,N$, is sensitive to the superradiant gain $G$ and the loss term $L$ that accounts for the decay of the matter-wave grating. In the mean-field BEC regime, the loss term $L$ is shown to track the thermal phase fluctuation as $L\propto a_s^{2/5}$ monotonically \cite{SR_dBEC}.

As the scattering length is reduced across the BEC-droplet crossover, our variational calculations show a decrease in the chemical potential (see Supplementary Material). On the BEC side, the enhanced superradiance efficiency is consistent with reduced thermal phase fluctuations. Within the phase-fluctuation theory for elongated Thomas-Fermi condensates, the phase-fluctuation parameter decreases with decreasing chemical potential at fixed temperature, atom number, and axial confinement \cite{Petrov01}. This framework, however, cannot be directly extrapolated to the droplet regime. For dipoles aligned with the long axis, previous studies have identified anti-roton behavior, characterized by a rapidly rising axial excitation branch and a redistribution of density-response weight among excitation branches \cite{Pal.2022,Houwman2024}. This behavior reflects the increasingly repulsive effective interaction at larger axial momenta \cite{Pal.2022}. Experimentally, we observe a maximum in superradiance efficiency near the crossover and a decrease further into the droplet regime (Fig.~\ref{fig1}(b)). The latter may involve changes in coherence, density, geometry, and excitation spectrum.

The superradiance process also allows to explore BEC-droplet phases via controlled depletion of the sample. During the scattering process, the initial atomic cloud (mother cloud) becomes depleted, forming two additional clouds (daughter clouds), controllable via the scattering rate (i.e.\ intensity) of the pump beam. By monitoring the expansion dynamics of the depleted mother cloud, including the aspect ratio measurement, one can map out the BEC-droplet crossover as a function of atom number as described in Fig.~\ref{fig1}(c).

\begin{figure*}[ht]
	\includegraphics[width=0.95\linewidth]{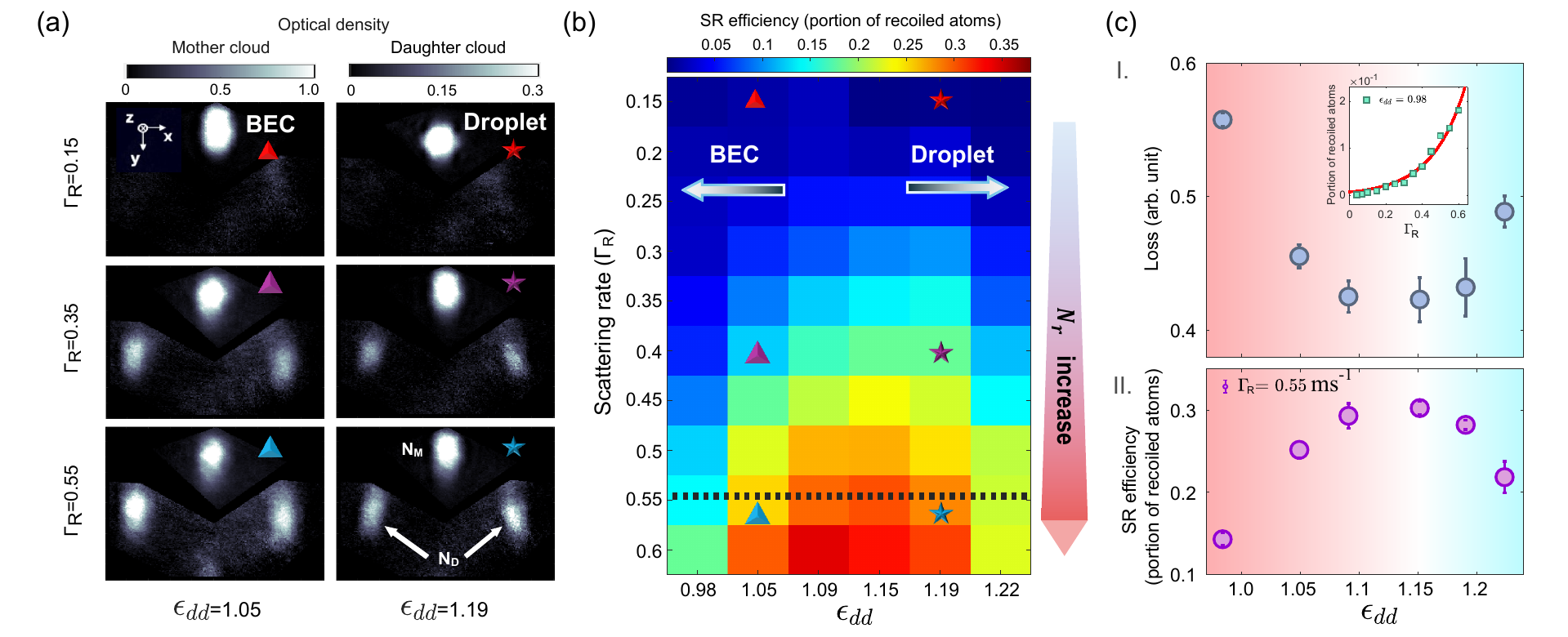}
	\caption{{\bf Superradiance threshold behavior from BEC to macrodroplet.} (a) Vertical absorption images of atomic momentum distribution after exposure to an off-resonant laser pulse of variable scattering rate. Images are taken after 10 ms time of flight. The scattering rates are 0.15, 0.35, and 0.55 ms$^{-1}$. Images on the left side represent the BEC regime ($\epsilon_{dd}=1.05$, prisms), whereas on the right side is the droplet regime ($\epsilon_{dd}=1.19$, stars). In both BEC and droplet phases, an increase in the scattering rate leads to a strong depletion of the mother cloud and thus larger daughter clouds. To clearly distinguish the aspect ratio change and increase of SR efficiency for larger $\Gamma_R$, different color scales are applied to mother and daughter clouds. (b) The portion of recoiled atoms versus the Rayleigh scattering rate, $\Gamma_R$, and dipolar strength, $\epsilon_{dd}$. The superradiance efficiency is maximum across the crossover and is progressively suppressed as the system moves away from the transition point. The absorption images corresponding to prisms and stars are depicted in (a). (c) An exponential increase in total number of recoil atoms is observed. The increase is proportional to the loss term, $L$, and is illustrated as a function of $\epsilon_{dd}$. The loss term reduces to a minimum value near the transition point and enhances again with a further increase of $\epsilon_{dd}$. Inset shows portion of recoiled atoms at $\epsilon_{dd} = 0.98$ fitted to the equation $N=A~\text{exp}\left({(\Gamma_R-L)t_{\text{pulse}}}\right)$ where $t_{\text{pulse}}$=100 $\mu$s and $A$ is \rev{an} arbitrary coefficient. In the fitting, $\Gamma_R$ is an independent variable (from 0.04 to 0.6 ms$^{-1}$). Lower panel: the portion of recoiled atoms for $\Gamma_R=0.55$ ms$^{-1}$ across the transition point, showing \rev{that the} loss term and superradiant efficiency $\left(\frac{N_D}{N_D+N_M}\right)$ are inversely related.}
	\label{fig2}
\end{figure*}

\paragraph*{\bf Superradiance threshold behavior across the BEC-droplet crossover}

We begin with a BEC of $^{166}$Er, containing $\sim$1.5-3.5(2)$\times 10^4$ atoms in the $\ket{m_j=-6}$ state \cite{ErMOT20, ErBEC23, SR_dBEC}. These atoms are trapped within a crossed optical dipole trap (ODT), which is composed of a pair of 1064 nm laser beams extending along the x and y axes with a beam waist of $\omega_{y-z}$=20~$\mu m$ and $\omega_{x-z}$=45~$\mu m$, respectively. Next, we adiabatically adjust the trap geometry over a period of 50~ms to form a quasi-1D trap. The trap frequency is then set to $(\omega_x,\omega_y,\omega_z)=2\pi\times(21, 290, 290)~$Hz. $\epsilon_{dd}=a_{dd}/a_s$ is controlled between 0.89 and approximately 1.4 where $a_{dd}=\frac{\mu_0\mu^2 m}{12 \pi \hbar^2}$ and $a_s$ ($\mu \approx 7\mu_B$ denotes magnetic dipole moment of erbium atoms) are dipolar and scattering length, respectively. To control the scattering length $a_s$, the magnetic field firstly changes from 1.4 G to 0.4 G gradually within 50 ms, together with direction change from z-axis to x-axis along elongated ODT direction (i.e.\ $\theta$=0). Then within 3 ms, the magnetic field is changed from 0.4 G to the target value right before the pump beam is switched on, while the field direction is kept along the x-axis (i.e.\ $\theta$=0).

 A transition from the BEC to the macrodroplet is realized by controlling the scattering length $a_s$, resulting in tunable $\epsilon_{dd}=a_{dd}/a_s$ where $a_{dd}$ is the dipolar length. In the first set of experiments, we measure the portion of recoiled atoms as a function of the Rayleigh scattering rate across the BEC-droplet transition when the atom number of the mother cloud is kept constant for variable $\epsilon_{dd}$. As shown in Fig.~\ref{fig2}(a, b), the recoiled atom number reaches a maximum near the BEC-droplet crossover, suggesting that the superradiance efficiency reveals the crossover. The increased superradiance efficiency in the BEC regime can be attributed to decreasing thermal phase fluctuations \cite{Petrov01, PhaseFlucExp}. These phase fluctuations contribute to the fitted loss term $L$ obtained from the recoiled atoms for variable $\Gamma_R$ (see inset of Fig.~\ref{fig2}(c)). However, in the quantum droplet regime, spatial coherence is affected by the dipolar and BMF interactions, influencing the chemical potential, and excitation energy \cite{Baillie2017, Pal.2022, Houwman2024}. Fig.~\ref{fig2}(b) shows the portion of recoiled atoms for varying $\epsilon_{dd}$. At a \rev{sufficiently} large scattering rate \rev{(e.g.\ $\Gamma_R = 0.55$ ms$^{-1}$)}, the superradiant efficiency shows non-monotonic behavior \rev{as a function of $\epsilon_{dd}$} (dashed line, see also Fig~\ref{fig2}(c)), being consistent with the expansion velocity, which reveals the property of the quantum state before the superradiant light is switched on (see the expansion velocity in Fig.~\ref{fig4}(a)).

\begin{figure*}[ht]
	\includegraphics[width=0.95\linewidth]{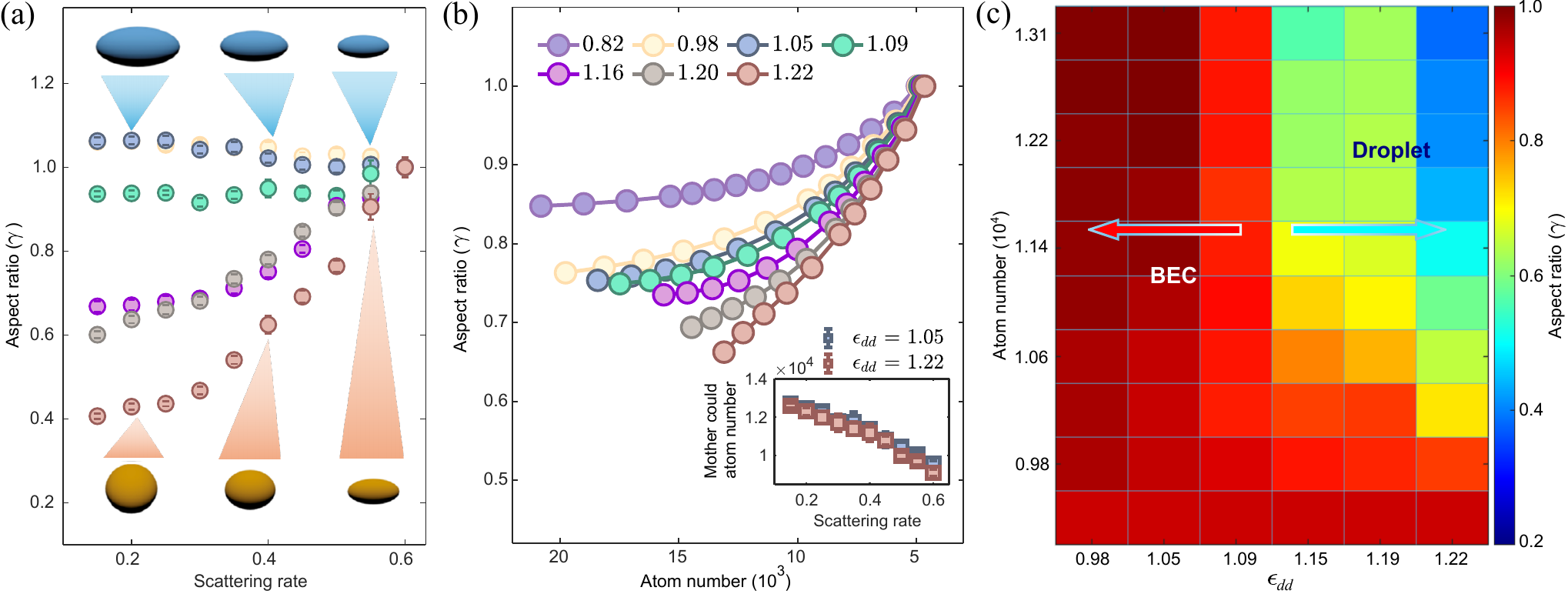}
	\caption{{\bf Dynamical response of the depleted mother cloud.} (a) The scaled aspect ratio ($\gamma$) of the mother cloud is shown as a function of scattering rate, $\Gamma_R$ for different values of $\epsilon_{dd} = 0.98,~1.05,~1.09$ in BEC regime and $\epsilon_{dd} =~1.15,~1.19,~1.22$ in droplet regime (from top to bottom curves). The aspect ratio is rescaled by the value corresponding to the largest scattering rate for each $\epsilon_{dd}$. Ellipsoids on top and bottom depict a schematic representation of the shape of the mother cloud at different scattering rates, $\Gamma_R$ for the BEC (upper, $\epsilon_{dd}=1.05$) and droplet (lower, $\epsilon_{dd}=1.22$) phases, illustrating distinct behavior of $\gamma$ in both regimes. (b) Theoretical simulation of aspect ratio (normalized) after 10 ms of free expansion. Inset shows the loss of mother cloud atom number with an increase of scattering rate, $\Gamma_R$. (c) Dynamical response of the depleted mother cloud is measured for the BEC-droplet crossover in $\epsilon_{dd}$-$N$ space.}
	\label{fig3}
\end{figure*}

\paragraph*{\bf Expansion dynamics following superradiant depletion}

We next analyze the expansion of the mother cloud following the superradiant pulse. The mean remaining atom number varies reproducibly with the scattering rate (and hence the pump intensity), while the short depletion time limits three-body loss during the pulse. This protocol provides rapid tuning of the mean atom number.

In Fig.~\ref{fig3}(a), a sample is first adiabatically prepared between the BEC (e.g.\ $\epsilon_{dd}$=0.98) and droplet (e.g.\ $\epsilon_{dd}$=1.22) regimes. Utilizing this light-induced depletion, we subsequently measure the expansion dynamics of the atomic cloud depleted by the superradiance process (i.e.\ mother cloud) and monitor the variation in the aspect ratio of the mother cloud after superradiant scattering. The first three curves shown in Fig.~\ref{fig3}(a) represent the aspect ratio for three $\epsilon_{dd}$ values of $0.98,~1.04,~1.08$ respectively, for a dilute BEC. \rev{Throughout, the aspect ratio is normalized by its value at the largest scattering rate for each $\epsilon_{dd}$, such that the curves converge to unity at high depletion and their relative trend directly reveals the phase-dependent response.} Notably, the aspect ratio of the mother cloud of the BECs remains almost constant with increasing scattering rate, as the droplet requires a minimum critical number of atoms and $\epsilon_{dd}$. In the BEC regime, a decrease in atom number slightly weakens the release energy (see supplementary note), causing a minor decrease in the aspect ratio. In the droplet regime at $\epsilon_{dd}$=1.22, the depletion in the mother cloud leads to a rapid increase in aspect ratio following the superradiance process, as shown in Fig.~\ref{fig3}(a). As the scattering rate increases, the aspect ratio of the mother cloud gradually increases for $\epsilon_{dd}\geq1.15$. The increase (decrease) of the aspect ratio in the droplet (BEC) regime is attributed to the release energy \cite{RE1, RE2}. Moreover, the initially small aspect ratio at a low scattering rate results from attractive interactions dominating over repulsive ones \cite{DDI_AR1, DDI_AR2, DDI_AR3}.

To verify this distinct expansion dynamics of the depleted sample after superradiance, we numerically calculate the aspect ratio after 10 ms of free expansion by solving the extended Gross-Pitaevskii equation (eGPE) with time-dependent variational Gaussian ansatz \cite{Santos16a, QD16_FF}. We solve for variational parameters, atom number, and condensate size (see supplementary note), which reflects the change in atom number for variable $\Gamma_R$ as shown in Fig.~\ref{fig3}(b) inset. The observed expansion dynamics, characterized by the aspect ratio of the sample, are in good agreement with Fig.~\ref{fig3}(b).

Next, using this approach, we can investigate how the depleted mother cloud responds in $\epsilon_{dd}-N$ space by analyzing the aspect ratio of the expanded cloud, with variable parameters $a_s$ and $\Gamma_R$. Importantly, our light-induced depletion method circumvents complications from three-body loss, enabling precise control over atom number in the diagram. In Fig.~\ref{fig3}(c), the aspect ratio of the sample exhibits a distinct transition between the BEC (towards the red arrow) and the droplet phases (towards the blue arrow) across this parameter space. Based on numerical calculations, the droplet phase cannot form below $\epsilon_{dd}\sim$1.1, even with large atom numbers ($N$), which is consistent with our experimental observations.

\paragraph*{\bf Effect of tilted dipoles on crossover signatures}

Finally, we employ two complementary approaches utilizing superradiant scattering, as illustrated in Figs.~\ref{fig2} and \ref{fig3}, to examine the BEC-droplet transition in the unprecedented regime. For example, we evaluate how tilted dipoles influence droplet formation within an elongated trap. We maintain a consistent Rayleigh scattering rate, thereby ensuring a stable superradiance gain term, to assess superradiance efficiency throughout the BEC to macrodroplet transition. Our measurements involve dipole orientations of $\theta=0^o$, $20^o$, and $30^o$ relative to the elongated axis. Fig.~\ref{fig4}(a) demonstrates that the recoiled atom fraction reaches its peak at $\epsilon_{dd}\sim 1.14$ for 0 degrees, with subsequent shifts to approximately $1.21$ and $1.28$ for 20 and 30 degrees, respectively. This indicates a shift in the BEC-droplet transition as $\theta$ increases. We confirm this observation by measuring the expansion dynamics across the crossover for various scattering lengths and dipolar orientations.

\begin{figure}[ht]
    \includegraphics[width=1\linewidth]{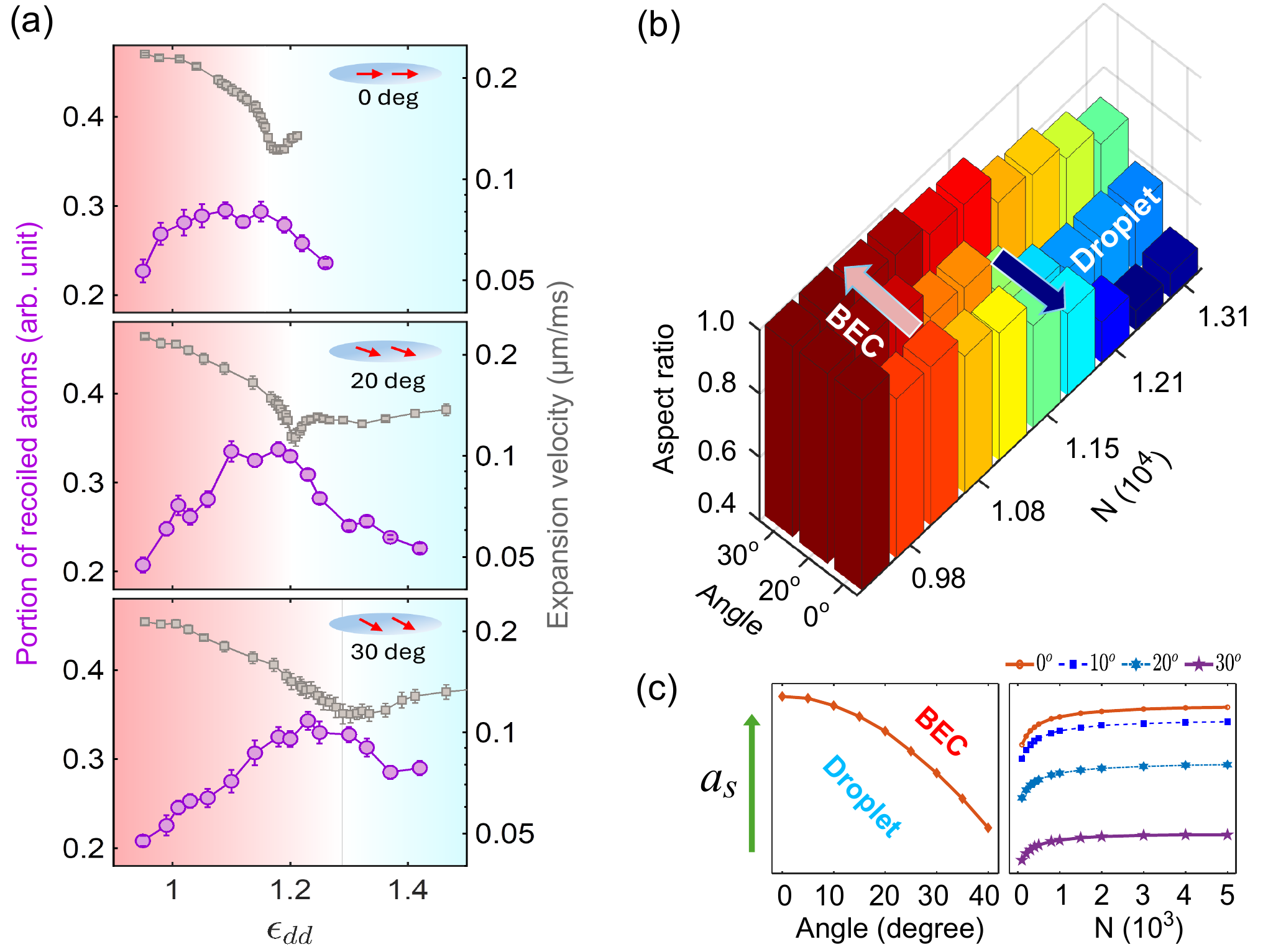}
    \caption{{\bf  Shift of crossover signatures with magnetic-
field orientation..} (a) The expansion velocity of the cloud is measured for different $\epsilon_{dd}$ values at magnetic field angles of 0°, 20°, and 30°. A clear shift in the point of lowest expansion rate is observed for all three angles (gray curves). The portion of recoiled atoms is measured for different $\epsilon_{dd}$ values from the BEC-droplet phase at magnetic field angles of 0°, 20°, and 30° (purple curves). (b) Experimental data for aspect ratio (normalized) of mother cloud in $\theta-$N space at $\epsilon_{dd}=1.22$. (c) Theoretical phase-diagram of BEC-droplet crossover in $a_s-\theta$ space for fixed atom number (left) and $a_s$-$N$ space for different angles (right).}
	\label{fig4}
\end{figure}

The histograms in Fig.~\ref{fig4}(b) represent the aspect ratio for $\epsilon_{dd}=1.22$ as a function of the dipolar angle $\theta$ and the number of atoms $N$ remaining in the mother cloud. The alignment of dipoles away from the elongated axis reduces the attractive dipolar energy. Thus, it requires lowering the scattering length to achieve a self-bound system, as illustrated in Fig.~\ref{fig4}(c) theoretically. In Fig.~\ref{fig4}(c), we plot the scattering length as a function of dipolar orientation and atom number where release energy becomes zero. The transition shifts to lower values of $a_s$ with increasing dipolar angles, whereas it saturates for large atom numbers regardless of dipolar orientation. The system possesses a self-bound property for negative values of the release energy or chemical potential. Therefore, their zero value corresponds to the minimum expansion velocity.

In conclusion, we have demonstrated that superradiant scattering effectively reveals quantum state properties in dipolar $^{166}$Er gas systems. Using this process as both probe and control, we have observed non-monotonic efficiency across the BEC-droplet transition. Through controlled atom depletion and analysis of expansion dynamics, we have explored this transition and mapped its shifting points under varying magnetic field orientations.

It is worth noting that the superradiant pulse transfers a controlled fraction of atoms into well-separated momentum states within $\sim$100~$\mu$s---essentially instantaneously compared with the trap period, after which the remaining mother cloud evolves from a well-defined initial condition. The final atom number is set deterministically by the pump intensity rather than by a stochastic decay process, which is precisely what enables us to characterize the superradiance response across the BEC-droplet crossover in the $\epsilon_{dd}-N$ space (see Fig.~\ref{fig3}(c)). This advantage may open a new opportunity over the sample, depleted with three-body recombination process, where it removes atoms preferentially from the high-density core, deposits recombination energy into the sample, and proceeds continuously over tens of milliseconds, so that the system evolves through a sequence of heated, non-equilibrium states whose atom number and temperature are entangled. 

Our findings pave the way for further investigation into quantum droplets and supersolids \cite{ss1, ss2, ss3} as well as light-matter interactions in dipolar systems. The light scattering technique we demonstrated offers a novel approach to probing responses influenced by phase coherence in low-dimensional dipolar systems, particularly in one- \cite{Kao2021} and two-dimensional \cite{He2025, zhen2025breaking,he2025a,he2026a} configurations. This measurement could provide insights into low-energy excitations in the droplet \cite{Baillie2017, Pal.2022, Houwman2024}. \rev{An intriguing extension is to analyze the statistical fluctuations of the superradiantly emitted light itself. Since the superradiant gain is seeded by density and phase fluctuations of the gas, shot-to-shot statistics of the emitted pulse could serve as a fluctuation spectroscopy of the crossover, complementing the mean efficiency measured here.} Additionally, harnessing the superradiant process offers a promising method to control atom number in the sample. Since the rapid superradiance-induced depletion process prevents three-body loss, it creates a unique opportunity to study droplet formation dynamics.

\section{Appendix}

\paragraph*{\bf Superradiant scattering and detection}
There is a three-body loss of atoms in the optical trap, with higher losses observed for lower $a_s$. Before pulsing a superradiance light, we control the atom number by adjusting the evaporation time ensuring that the total atom number is the same for all $a_s$ across the BEC-droplet crossover. To probe a quantum state of the sample with superradiant Rayleigh scattering, we shine a single off-resonant 583~nm pump beam on an elongated sample. The pump beam, which is linearly polarized along the z-direction (as shown in Fig.~\ref{fig1}), is directed at an angle perpendicular to the elongated axis and has a beam waist $\sim$100 times larger than the sample. \rev{The pulse duration is fixed at $t_{\text{pulse}}=100~\mu$s for all measurements.} To reduce the effect of the external trap, the all-optical dipole traps (ODTs) are turned off 60~$\mu s$ before the pump beam is on\rev{; owing to the short pulse duration, turning off the trap earlier or later within this window has no measurable effect on the recoiled fraction}. We tune the Rayleigh scattering rate ($\Gamma_R$) up to 0.6 ms$^{-1}$ for different magnetic fields, based on independent calibration (see the supplementary information of \cite{SR_dBEC}). The change in $\Gamma_R$ effectively modulates the gain term $G$ of the superradiance.

Following collective light scattering, we record the atomic momentum distribution after a 10 ms time-of-flight free expansion followed by absorption imaging with fringe removal \cite{Song2020}, using the $\sigma^{-}$ circularly polarized 401~nm broad-linewidth transition ($4f^{12}6s^2({}^3H_6)\rightarrow 4f^{12}({}^3H_6)6s6p({}^1P_1)$) directed along the z-axis. The total number of recoiled atoms in the daughter cloud $N_D$ is equal to the sum of atoms in both the left and right clouds, as shown in Figs.~\ref{fig1} and \ref{fig2}(b).

\paragraph*{\bf Theoretical model}
The expansion dynamics of the depleted mother cloud are simulated by solving the extended Gross-Pitaevskii equation, including the Lee-Huang-Yang correction, with a time-dependent Gaussian variational ansatz \cite{Santos16a, QD16_FF}. The equations of motion for the variational widths are integrated for the measured initial atom numbers and trap parameters, and the aspect ratio is evaluated after 10~ms of free expansion. Details of the variational equations and the extraction of the loss term $L$ from the recoiled-atom fraction are provided in the Supplementary Information.

\vspace{10pt}
\paragraph*{\bf Data Availability} The data that support the findings of this work are available upon request.

\vspace{10pt}
\paragraph*{\bf Acknowledgement}
This work was supported by the RGC through 16306321, 16306922, 16302123, 16305024, C6009-20G and RFS2122-6S04. GBJ acknowledges support from the Gordon and Betty Moore Foundation, grant DOI 10.37807/GBMF13794.

\bibliography{allref}

\end{document}